\documentstyle[aps,prd,preprint,floats,epsfig,tighten]{revtex}

\begin{document}
\draft
\title{A Generalised Area Law for Hadronic String Reinteractions\thanks{Work 
supported by  the Swedish Natural Science Research Council,
contract F--PD 11264--301 and the U.S. Department of Energy, 
contract DE--AC03--76SF00515.}}
\author{Johan Rathsman}
\address{Stanford Linear Accelerator Center \\
Stanford University, Stanford, California 94309, USA\\
e-mail: rathsman@slac.stanford.edu}
\preprint{ \vbox{\flushright SLAC--PUB--8034--Rev \\ March, 1998 }}
\maketitle
\begin{abstract}
A new model for hadronic string reinteractions based on a 
generalised area law is presented. The model describes
both the hadronic final states in $e^+e^-$ annihilation
and the diffractive structure function in deep inelastic scattering.
The model also predicts a shift in the W-mass reconstructed from 
hadronic decays of W-pairs of the order 65 MeV. 
\end{abstract}

\newpage

The problem addressed in this letter is how to translate a partonic final state
consisting of quark and gluons, calculated in perturbative quantum chromo
dynamics (QCD), into a final state of hadrons. Since this a process that takes
place at low momentum transfer, perturbative methods cannot readily be
applied.  Instead one has to take resort to  phenomenological models, like the
Lund string model \cite{lundstring},  to describe this  non-perturbative
transition.  The string model is a semi-classical model  which, when combined
with perturbatively calculated partonic cross-sections,  gives a good overall
description of observed hadronic final states,  especially in $e^+e^-$
collisions. 

One of the basic assumptions of the Lund string model is that the colour field
between two colour charges forms a flux-tube with the dynamical properties of
a one-dimensional relativistic string. In the simplest case with just one quark 
and one anti-quark connected by a string this gives the so called yo-yo model.
Another assumption which is usually 
made in the application of the Lund string model is that two 
different strings will hadronise independently of each other even if they
overlap. In other
words there is no cross-talk or string reinteractions 
in a system consisting of more than one string. 

This letter presents a new general model for taking string reinteractions into
account in hadronic final states.  String reinteractions as a way of
understanding diffractive events was suggested a long time ago \cite{artru}.
Similar ideas have also been considered in connection with W-pair production
\cite{gpz,skprl,wrec1,wrec2,wrec3,wrec4,wrec5,wrec6,wrec7} 
and in the model for soft colour interactions (SCI) which originally
was formulated for rapidity gaps in deep inelastic scattering \cite{sci1,sci2}. 

The starting point for the present model will
be the string configuration given by the colour flow in a reaction on  partonic
level where the parton configuration has been calculated perturbatively. On top
of that interactions in the form of string rearrangements will be added. Thus,
giving an altered colour configuration and thereby a different  hadronic final
state. The main difference compared to SCI is that the model presented here
considers string reinteractions where the colour fields in the two  strings
interact whereas in SCI the perturbative partons interact with a hadronic 
background field.  

{\bf The model.} 
The Lund String model is based on the so called area law \cite{wilson}.
Simply put the area law means that configurations with a large area are 
exponentially suppressed. Within the Lund String model the probability
for a configuration with area $A$ is given by 
 $P \propto \exp(-bA)$
where $b$ is a phenomenological parameter of the order $0.6$ GeV$^{-2}$
if the area is calculated in energy-momentum coordinates. In the following,
the area for a piece of string spanned between two partons
$p_i$ and $p_j$ will be calculated as
$A_{ij} = (p_i+p_j)^2-(m_i+m_j)^2 = 2(p_ip_j-m_im_j)$
such that the area vanishes for two massive partons at rest. This way of 
defining the area reduces to the ordinary 
$A=(E_i+p_{z,i})(E_j-p_{z,j})=(E+|p_z|)^2$, in the
center of mass system with the partons along the $z$-axis, 
when the quark masses are neglected.

\begin{figure}
\begin{center}
\mbox{\epsfig{figure=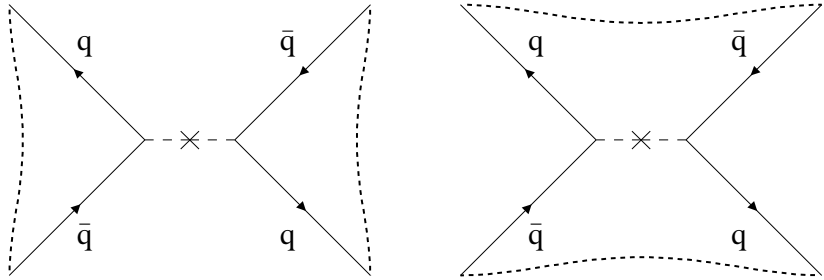,width=12cm}}
\end{center}
\caption[*]{Illustration of two different string configurations
in W-pair production in the lowest order 
(the strings are indicated with dashed lines). }
\label{fig:wpic}
\end{figure}

For a single string the only way to decrease the area is by 
"popping" quark pairs from the vacuum. However, for a system consisting
of several strings it may also be possible to decrease the area by
doing a string rearrangement. The simplest example is given
by the situation where there are two strings, each consisting of
a quark anti-quark pair as illustrated in Fig.~\ref{fig:wpic}. Labeling
the momenta of the partons in the two initial strings
($p_1$, $p_2$) and ($p_3$, $p_4$) respectively 
it may be possible that the system can decrease its area by making a 
string rearrangement into $(p_1,p_4)$ and $(p_2,p_3)$. 
The initial area is given by $A^{old}=A_{12}+A_{34}$ 
where as the alternative configuration has the area 
$A^{new}=A_{14}+A_{23}$.

Generalising the area law to be applicable to a system of strings 
the probabilities for these two different configurations are 
given by $\exp(-bA^{old})$ and $\exp(-bA^{new})$
respectively times some overall factor depending on the parton momenta
which is the same for both of them. 
In this paper it will be assumed that the probability for
a string rearrangement is proportional to the normalised difference 
between the two configurations
\begin{equation}
 P \propto 
 \frac{\exp(-bA^{new})-\exp(-bA^{old})}{\exp(-bA^{new})}
 = 1-\exp(-b\Delta A)
\end{equation}
where $\Delta A=A^{old}-A^{new}$ is the area difference 
between the old and the new configuration. 

In addition there should also be a colour suppression factor of the order 
$\frac{1}{N_C^2}$
reflecting the required colour matching of the two strings.
Since in this 
model it is the strings that are thought to interact it is the colour field
in the string that is relevant and not the colour charges at the
endpoints of the string. This should be compared with the fact that
in a quark--gluon--anti-quark string the quark and the anti-quark are not in
a colour singlet state. The string carries an octet charge and thus the 
colour factor required for matching the colour field in the two strings
is of the order $\frac{1}{N_C^2}$.

The phenomenological ansatz of the model is to have a string
rearrangement between a pair of string pieces
with the probability 
\begin{equation}\label{prob}
 P  = R_0\left[1-\exp(-b\Delta A)\right]
\end{equation}
where $R_0$ is a nonperturbative parameter of the order 
$\frac{1}{N_C^2}$. The parameter will be fixed to $R_0=0.1$ 
by comparing
with data on rapidity gap events in deep inelastic scattering.
This formula for the probability should be compared with the SCI model
where the probability is assumed to be a constant, i.e. without the
suppression factor 
$\left[1-\exp(-b\Delta A)\right]$\footnote{In an earlier version of 
the model the string-length was used to formulate a suppression factor.
I would like to thank Gunnar Ingelman for suggesting to use the string-area
instead when deriving the suppression factor.}. The actual 
implementation of the string rearrangements in the model is very similar to 
the one used for the SCI model. In short there is a loop over all
pairs of string pieces which have an alternative string configuration. 
A string rearrangement between the two string pieces 
is then made  with the probability given by Eq.~(\ref{prob}).
In principle one has to worry about the fact that the  order of the pairs 
in the loop can make a difference. However, since the probability
to make a string rearrangement is small, this is a minor problem
and the effects have been neglected.
The model is implemented in the Lund Monte Carlo (MC) 
framework and the code can be obtained from 
{\tt http://www3.tsl.uu.se/thep/rathsman/gal/}.

{\bf Electron-positron annihilation.} 
Since in this model the string rearrangement is thought of as
a general phenomenon and not connected with some hadronic background field
as in SCI \cite{sci1,sci2}
it has first of all to be retuned to data from
$e^+e^-$ annihilation at $\sqrt{s}=M_Z$. For this purpose the Jetset Monte 
Carlo version 7.4 \cite{jetset} will be used together with the model.
For easy reference most of the times 
the default version of Jetset will be used for comparison instead of data
and differences smaller than a few percent will be considered satisfactory.

\begin{figure}[tp]
\begin{center}
\mbox{\epsfig{figure=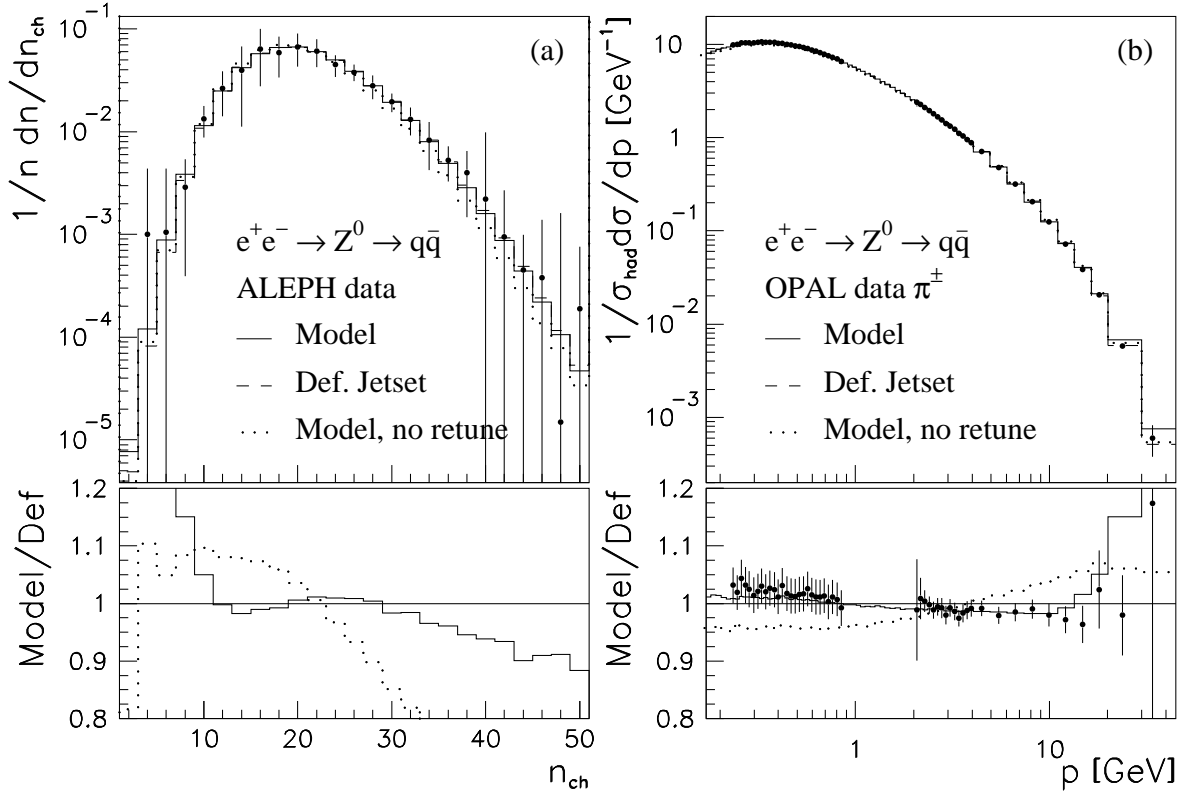,width=16cm}}
\end{center}
\caption[*]{The charged multiplicity distribution (a) and 
the momentum distribution for $\pi^\pm$ (b). The model (solid line) 
is compared with default Jetset (dashed line)
and the model without retuning (dotted line). The data are from the ALEPH
and OPAL collaborations respectively with the statistical and systematic errors
added in quadrature. The ratio of the model, with (solid) and without (dotted)
retuning, to default Jetset is shown below, (b) also shows the ratio of 
data to default Jetset. }
\label{fig:fig2}
\end{figure}

The model is tuned using the particle multiplicities 
and momentum distributions. By retuning the parameter $b$ in the fragmentation 
function to $b=0.45$ GeV$^{-2}$ and the cut-off in the parton showers to
$Q_0=2$ GeV, the mean multiplicity of charged particles ($<\!n_{ch}\!>$=20.9)
as well as the 
dispersion ($D=\sqrt{\!<\!n_{ch}^2\!>\! - \!<\!n_{ch}\!>\!^2}=6.2$)
are the same as in the default version. The resulting 
charged multiplicity distribution is shown in
Fig.~\ref{fig:fig2}(a) together with the default version of Jetset and
data from the ALEPH collaboration \cite{alephdata}.
The multiplicities for individual mesons and baryons are typically within a 
percent of the default version. 

Considering the momentum 
distributions the differences are typically of the order a few percent 
which is about the size of the errors in the data.
As an example Fig.~\ref{fig:fig2}(b) shows the $\pi^\pm$ momentum 
distribution for the new model with and without retuning compared to
default Jetset and also compared with data from the OPAL collaboration
\cite{opalcharged}. As can be seen from the figure the new model is actually 
closer to the data than the default version and it should 
be possible to get a good agreement with data after a dedicated retuning.

With the charged multiplicity and the momentum distributions used
to retune the model, the string effect and the rapidity gap distribution will
be used to check if the model is a viable alternative to the normal
string model as implemented in Jetset.
The string effect \cite{string_effect1,string_effect2,string_effect3} 
gives a measure of the colour structure
in three-jet events in $e^+e^-$ annihilation. In the the lowest order 
diagram $e^+e^- \rightarrow q\bar{q}g$ the string goes from the quark
via the gluon to the anti-quark. In turn this means that there will be
a depletion of particles produced between the two quark jets compared to 
the particle production between the gluon jet and either of the quark jets
since there is no string between the quarks. 

\begin{figure}[tp]
\begin{center}
\mbox{\epsfig{figure=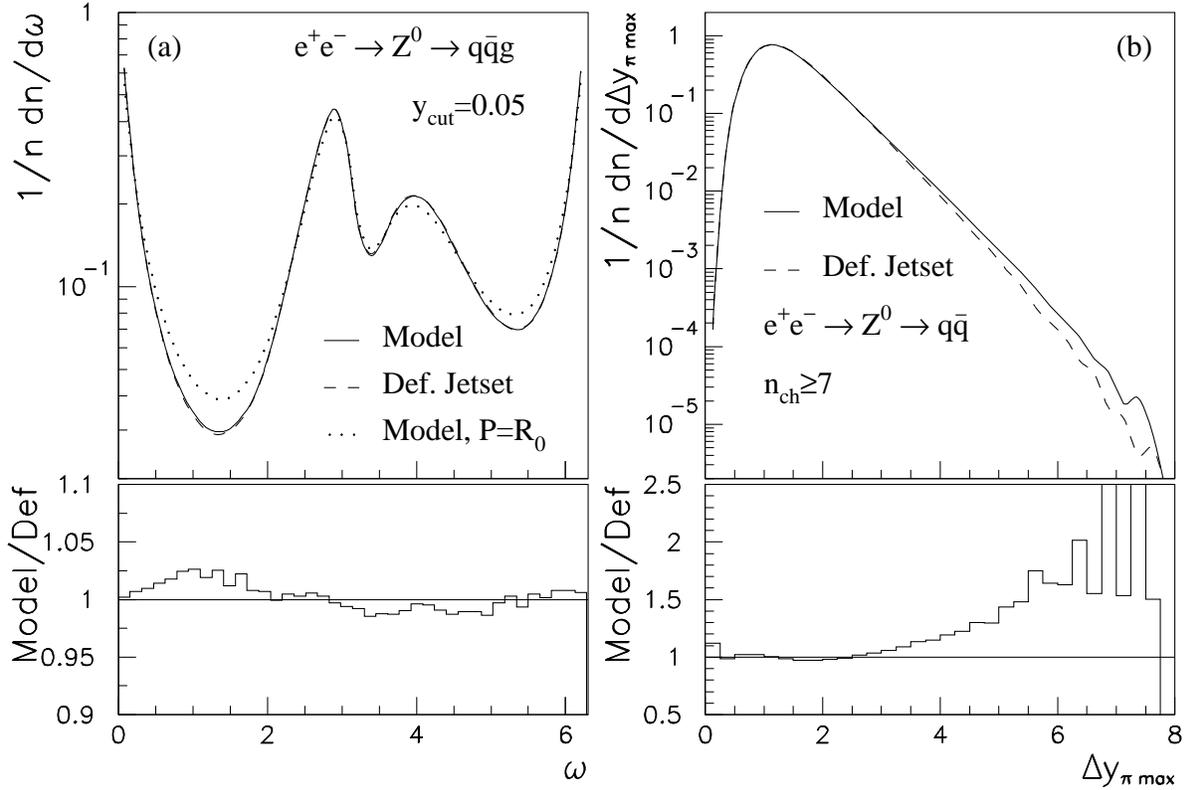,width=16cm}}
\end{center}
\caption[*]{(a) Illustration of the string effect as explained in the text
and (b) the largest rapidity gap distribution.
The model (solid line) is compared with default Jetset (dashed line)
and in (a) the model without the  area suppression factor (dotted line).
The ratio of the model and default Jetset is shown below. }
\label{fig:fig3}
\end{figure}

To study the string effect the JADE algorithm was used
for jet reconstruction with
the resolution $y_{cut}=0.05$. The events giving three jets
were analysed by projecting all particles on to the plane spanned by the
two most energetic jets. Fig.~\ref{fig:fig3}(a) shows the particle flow
in this plane as a function of the angle ($\omega$) 
from the most energetic jet
with $\omega$ defined such that the second jet has 
$\omega < \pi$. With the least energetic jet normally being the gluon
jet this shows the relative depletion between the two quark jets. As can be
seen from the figure the
difference between the default version of Jetset and 
the model is very small (below a few percent as shown by the lower
histogram giving the ratio of the two).
For illustration the model without the area suppression 
factor is also shown.

Another  observable which is  sensitive to the colour structure of an event is
the rapidity gap distribution. In $e^+e^-$ annihilation the rapidity is usually
defined  as  $y=\frac{1}{2}\ln\frac{E+p_z}{E-p_z}$, where the $p_z$ is the
momentum  along the thrust  axis and $E$ is the energy.  The rapidity gap
distribution has been measured by the SLD collaboration \cite{slddata} where
they assumed the mass to be the $\pi^{\pm}$ which will be denoted $y_{\pi}$ in
the following. Fig.~\ref{fig:fig3}(b) shows the distribution of the largest gap
in an  event $\Delta y_{\pi \max}$. As can be seen from the figure, the
difference between the model and the default version of Jetset is small for  
$\Delta y_{\pi \max}<4$  but for larger values the difference grows and becomes
as large as a factor 2 for $\Delta y_{\pi \max} > 6$.  However, judging by 
figure 1 in \cite{slddata} the errors in the data points in this region
are of the same size or even larger so at this point it is not possible to draw
any conclusions from the existing measurement\footnote{The SLD collaboration
has approximately three times more data on tape which is in the process of 
being analysed\cite{muller}.}.
The important point to notice is that the model does not give any plateau in
the rapidity gap distribution as one might expect but only  a gradual decrease
of the slope in the exponential suppression of large gaps.

{\bf Diffractive deep inelastic scattering.} 
As already mentioned the free parameter in the model $R_0=0.1$
is determined by calculating the diffractive structure function in deep
inelastic scattering and comparing with data from the H1 collaboration
\cite{h1data}.
The calculation is done by implementing the model in 
the Lepto Monte Carlo \cite{lepto} version 6.5
with the CTEQ 4 leading order parton distributions \cite{CTEQ4}. The only
changes with respect to the default version of Lepto is to use
the same values for the 
cut-offs in the initial and final state parton showers 
($Q_0=2$ GeV) and the 
hadronisation parameter ($b=0.45$ GeV$^{-2}$) as for $e^+e^-$ annihilation
given above\footnote{In addition version 2 of the sea-quark treatment
(see \cite{sci2})
was used with the width of the mean virtuality set to 0.44 GeV. However,
the result is not sensitive to this choice.}.
The diffractive structure function was then evaluated
using a subroutine from the HzTool package \cite{hztool}.

\begin{figure}[tp]
\begin{center}
\mbox{\epsfig{figure=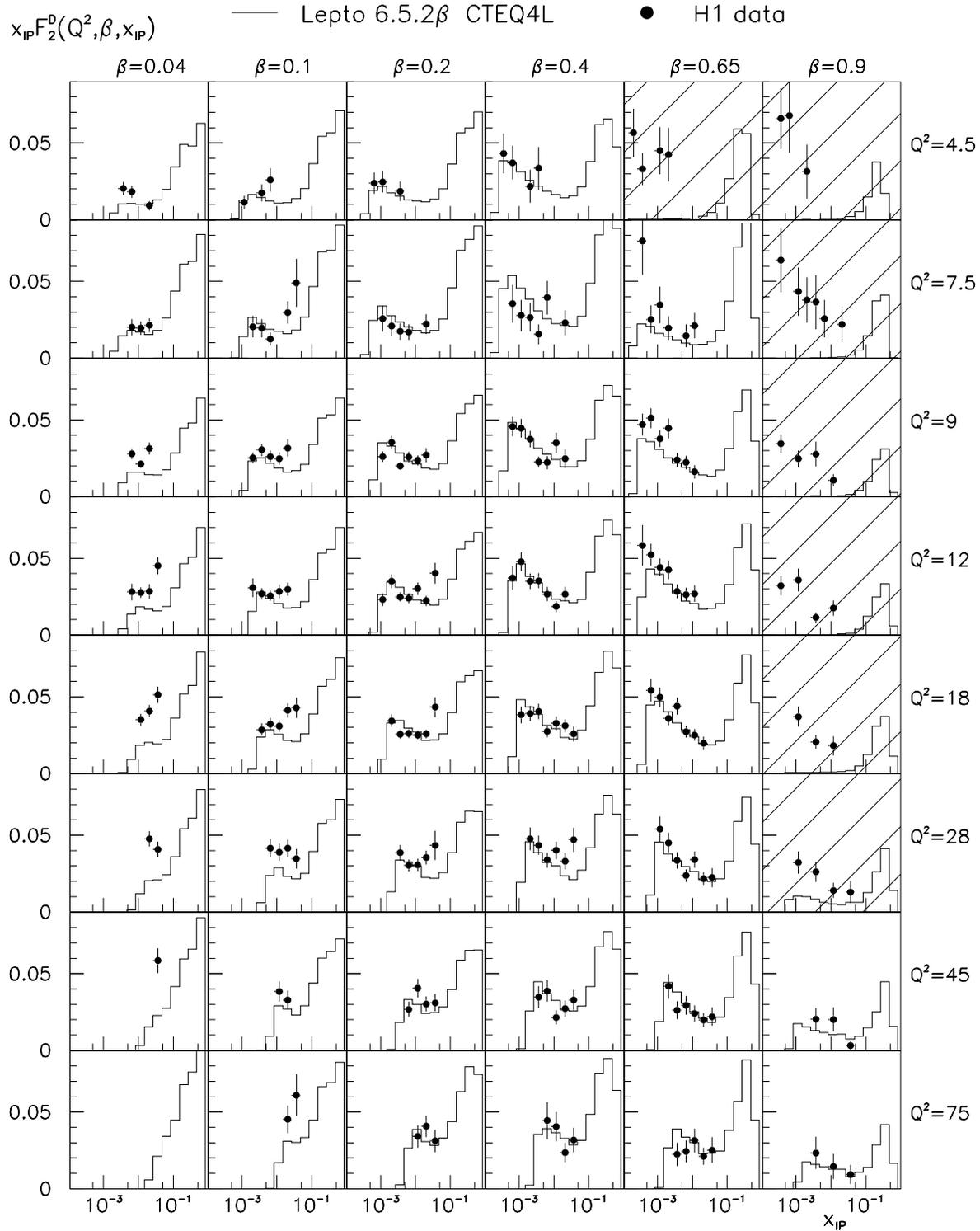,width=16cm}}
\end{center}
\caption[*]{The diffractive structure function obtained with the model
applied to Lepto compared to data from the H1 collaboration. The hashed
plots corresponds to kinematic points where the mass of the diffractive 
system $M_X$ is smaller than 2 GeV which is the cut-off in the matrix element.}
\label{fig:f2d3}
\end{figure}

The agreement between the resulting diffractive structure function 
calculated from the
model and the H1 data is very good as is shown in Fig.~\ref{fig:f2d3},
especially if one takes into account that there is
only one free parameter in the model. Both the so called Pomeron exchanges
which are thought to dominate for small 
$x_{\mbox{\tiny I\hspace{-1pt}P}}$, with $x_{\mbox{\tiny I\hspace{-1pt}P}}$
being the longitudinal momentum of the Pomeron with respect to the 
proton\footnote{With the mass of the 
diffractive system denoted by $M_X$, the photon virtuality $Q^2$ and W
being the mass of the hadronic system, 
$x_{\mbox{\tiny I\hspace{-1pt}P}}\simeq\frac{Q^2+M_X^2}{Q^2+W^2}$
and $\beta\simeq\frac{Q^2}{Q^2+M_X^2}$.},
as well as the other Regge exchanges which are important in the transition 
region $0.01<x_{\mbox{\tiny I\hspace{-1pt}P}}<0.1$ are explained by the model. 
The model only fails
for small masses of the diffractive system $M_X^2=Q^2\frac{1-\beta}{\beta}$
which are not included in the model 
because of the cut-off $M_X^2>4$ GeV$^2$ in the matrix-element. 

{\bf W-pair production.} 
The precise measurement of the W-mass is an important test of
electroweak theory. At LEP2 W-pairs are produced close to threshold in 
$e^+e^-$ annihilation.
One way of measuring the W-mass directly is to reconstruct it
from hadronic decays. When both of the W's decay hadronically 
one has to take into account the possibility that the two
hadronic systems interfere with each other\footnote{The separation,
at LEP2, between the two W's before they decay is small ($\sim 0.1$ fm) 
on a hadronic scale.}. 
This was first studied
in \cite{gpz} whereas the effects on the W-mass was first considered in
\cite{skprl}.  Later there have been several different models suggested 
for modeling the effects of colour reconnections 
on the reconstructed W-mass \cite{wrec1,wrec2,wrec3,wrec4,wrec5,wrec6,wrec7}.
For a recent comparison of
different models with data see e.g. \cite{watson,opalw}.

\begin{figure}[tp]
\begin{center}
\mbox{\epsfig{figure=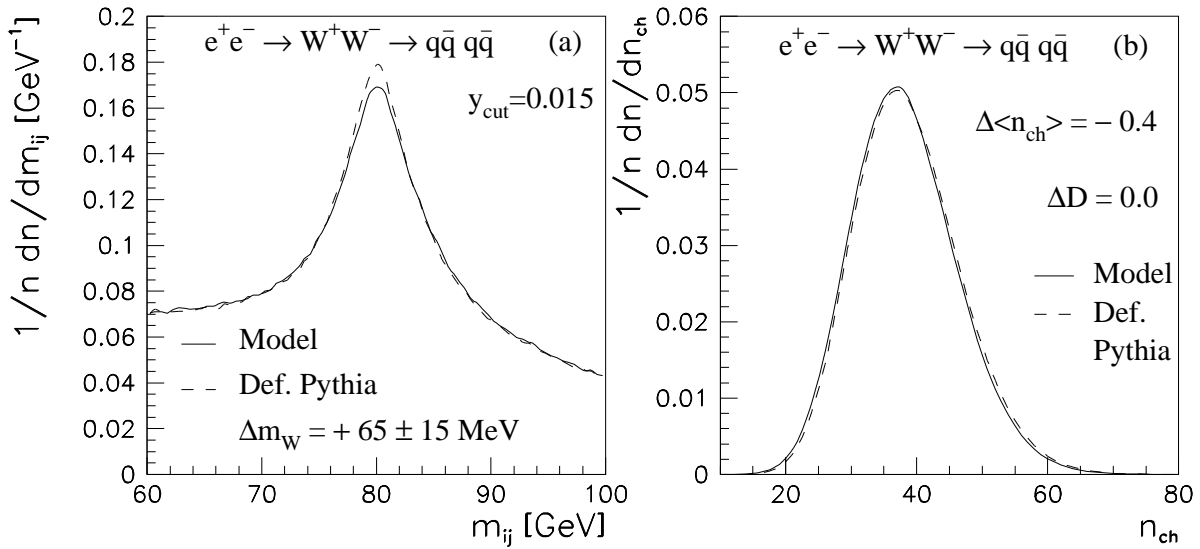,width=16cm}}
\end{center}
\caption[*]{The dijet mass spectrum (a) and the charged multiplicity (b)
for W-pairs decaying hadronically produced in $e^+e^-$ annihilation at 
$\sqrt{s}=183$ GeV.
The model (solid) is compared with default Pythia (dashed). As indicated,
the reconstructed mass is shifted with 
$\Delta m_W = m_W^{Model} - m_W^{Def} = 65\pm15$ MeV,
the mean multiplicity is shifted with 
$\Delta\!<\!n_{ch}\!> = <\!n_{ch}\!>\!^{Model} - \!<\!n_{ch}\!>\!^{Def} = - 0.4$ 
whereas the dispersion 
$D=\sqrt{\!<\!n_{ch}^2\!>\! - \!<\!n_{ch}\!>\!^2}$ is
unchanged. }
\label{fig:wfig}
\end{figure}

In the present model, which has been implemented using the Pythia Monte Carlo 
\cite{jetset} version 5.7, similar effects are expected. Fig.~\ref{fig:wfig}(a) 
shows the dijet mass spectrum for W-pairs (produced in $e^+e^-$ annihilation at 
$\sqrt{s}=183$ GeV) decaying hadronically using
the JADE algorithm for jet reconstruction with $y_{cut}=0.015$ and
considering only those events giving four jets. Fitting a Breit-Wigner
form plus a constant gives a reconstructed W-mass which is $65\pm15$ MeV
larger in the model than in the default Pythia version 
(the error is statistical).
This gives an estimate of how large a mass shift one can 
expect even though the precise number will depend on the analysis method used.

The obtained mass-shift is approximately twice as big as 
the final statistical error for LEP2 predicted in \cite{lep2ws}. 
On the one hand this
could ruin the usefulness of the hadronic decays for an exact determination
of the W-mass but on the other hand it could be used as a probe for 
non-perturbative dynamics. Other observables that have been used for
comparing different models with data are the multiplicity and 
thrust distributions (see e.g. \cite{watson,opalw}). 
The multiplicity distribution in the present model is shown 
in Fig.~\ref{fig:wfig}(b). Compared with default Pythia, the
mean multiplicity is somewhat smaller, $\Delta\!<\!n_{ch}\!> = - 0.4$, 
whereas the dispersion is unchanged.
The latter is also true for the thrust distribution.

{\bf Conclusions.} 
A new general model for string rearrangements in hadronic final states
has been presented. For hadronic final states in $e^+e^-$ annihilation
the model gives small differences compared 
to the Jetset Monte Carlo which in general describes data very well.
Thus the model is a viable extension of the ordinary Lund string model.
At the same time the model describes the diffractive structure function
in deep inelastic scattering. The model also predicts an enhancement for
large rapidity gaps in $e^+e^-$ annihilation which in principle should
be measurable with more statistics and a shift in the W-mass
reconstructed from W-pairs decaying hadronically. With more data the
model can be further tested and possibly
provide a probe into nonperturbative QCD phenomena.

{\bf Acknowledgments.}
I would like to thank Gunnar Ingelman and Anders Edin for useful
discussions on the modeling of non-perturbative effects in general,
and Gunnar Ingelman for valuable comments on the manuscript.
I would also like to thank Torbj\"orn Sj\"ostrand for helpful 
communications on the Lund string model and Hannes Jung for providing
the HzTool routine for calculating the diffractive structure function.

\end{document}